\documentstyle[12pt,epsf]{article}
\newcommand{\beq}{\begin{eqnarray}}
\newcommand{\eeq}{\end{eqnarray}}
\newcommand{\eq}{\ref}

\newcommand{\asms}{\alpha_{{}^{\overline{MS}}}}
\newcommand{\MSB}{\overline{MS}}
\newcommand{\MOM}{\widetilde{MOM}}

\newcommand{\gdue}{g^2_{\widetilde{MOM}}}
\newcommand{\Lam}{\Lambda_{\widetilde{MOM}}}
\newcommand{\Lams}{\Lambda_{\overline{MS}}}
\newcommand{\Lamlatt}{\Lambda_{{}^{LATT}}}
\newcommand{\Gev}{{\rm GeV}}
\newcommand{\Mev}{{\rm MeV}}
\newcommand{\ewxy}[2]{\setlength{\epsfxsize}{#2}\epsfbox[-40 60 640
590]{#1}}
\begin{document}
\setcounter{page}{1}
\begin{flushright}
IFUP-TH 23/96 \\
LPTHE 95/24 \\
Liverpool U.\ LTH 398\\
FERMILAB-PUB-97/169-T\\
Edinburgh 97/5\\
\end{flushright}
\centerline{\bf{\huge{$\alpha_s$ from the Non-perturbatively Renormalised}}}
\centerline{\bf{\huge{Lattice Three-gluon Vertex}}}
\vskip 0.8cm
\centerline{\bf{ B.\ All\'es$^{a,}$\footnote{Address after April 1997:
Dipartimento di Fisica, Sezione Teorica, Universit\`a Degli Studi di Milano,
Via Celoria 16, 20133-Milano, Italy.}, D.S.\ Henty$^{b}$,
H.\ Panagopoulos$^{c}$, C.\
Parrinello$^{d}$,
C.\ Pittori$^{e}$, D.G.\ Richards$^{f,b}$}} 
\centerline{$^{a}$ Dipartimento di Fisica
Universit\`a di Pisa}
 \centerline{Piazza Torricelli 2, 56126 Pisa, Italy}
\centerline{$^b$ Dept.\ of Physics \& Astronomy, University of Edinburgh,}
\centerline{Edinburgh EH9 3JZ, United Kingdom}
\centerline{(UKQCD Collaboration)}
\centerline{$^c$ Department of Natural Sciences, University of Cyprus}
\centerline{CY-1678 Nicosia, Cyprus}
\centerline{$^d$ Dept.\ of Mathematical Sciences, University of Liverpool}
\centerline{Liverpool L69 3BX, U.K.}
\centerline{(UKQCD Collaboration)}
\centerline{$^e$ L.P.T.H.E., Universit\'e de Paris Sud, Centre
d'Orsay,} \centerline{91405 Orsay, France.}
\centerline{$^f$ Fermilab, P.O.\ Box 500, Batavia, IL 60510, USA}
\begin{abstract}
We compute the running QCD coupling on the
lattice by evaluating two-point and three-point off-shell gluon
Green's functions in a fixed gauge and imposing
non-perturbative re\-nor\-ma\-li\-sa\-tion  con\-di\-tions on
them.
Our exploratory study is performed in the quenched approximation at 
$\beta=6.0$ on $16^4$ and $24^4$ lattices. We show that,
for momenta in the range $1.8-2.3 \ \Gev$, our coupling runs
 according to the two-loop asymptotic formula, allowing a
precise determination of the corresponding $\Lambda$ parameter. 
The role of lattice artifacts and finite-volume effects 
is carefully analysed and these appear to be under control in the momentum
range of interest.
Our renormalisation procedure corresponds to a momentum subtraction scheme 
in continuum field theory, and therefore lattice perturbation theory is not 
needed in order to match our results to the $\MSB~$
scheme, thus eliminating a major source of uncertainty 
in the determination of $\asms$. 
Our method can be applied directly to the unquenched case.

\end{abstract}
\vskip 0.8cm
\newpage
\section{Introduction}
\label{sec:intro}
The running coupling $\alpha_{s} (\mu)$, where $\mu$ is a momentum scale,
is a fundamental QCD quantity providing the link between low and
high-energy
properties of the theory. Given a renormalisation scheme, $\alpha_s (\mu)$
can be measured experimentally for a wide range of momenta.
A precise determination of $\alpha_s (\mu)$  (or equivalently of the scale
$\Lambda$ determining the rate at which $\alpha_s$ runs) is extremely
important
as it would fix the value of a fundamental parameter in the Standard Model,
providing bounds on new physics.

Computing $\alpha_s $ is a major challenge for the lattice
community. Several different lattice definitions of the renormalised
coupling have so far been investigated~\cite{Aida,Nara,Bali,Michael}.
Apart from the static quark potential
approach~\cite{Bali}, which involves a phenomenological
parametrisation of the interquark potential,
one feature of all other definitions of the coupling
is that lattice perturbation
theory (LPTH) has been used in order to 
convert the measured numerical value into a value for $\asms$. 
Despite recent proposals 
to improve the convergence of LPTH series~\cite{boost},
this step still provides one important source of systematic errors in the
final
prediction for $\asms$. At present, systematic errors, namely
quenching, discretisation effects, finite volume effects and LPTH,
dominate statistical ones for state-of-the-art computations. 

We investigate here a more recent proposal~\cite{io} for the
determination of $\alpha_s$ from the renormalised three-gluon vertex.
This is achieved by evaluating two-point and three-point off-shell
Green's functions of the gluon field on the lattice, in the Landau
gauge, and imposing non-perturbative renormalisation conditions on
them, for different values of the external momenta.  By varying the
renormalisation scale $\mu$, one can determine $\alpha_{s} (\mu)$ for
different momenta from a single simulation and analyse the
$\mu$-dependence of the coupling.  In particular, one can investigate
if the asymptotic behaviour is reproduced for large momenta. In
practice, for a given choice of the lattice parameters, one needs to
choose $\mu$ in a range of lattice momenta such that both finite
volume effects and discretisation errors are under control. If such a
momentum region exists and if the coupling is found to run according
to the two-loop asymptotic formula, then we can get a meaningful
measurement of $\alpha_{s} (\mu)$ in our renormalisation scheme which
can then be related perturbatively to other definitions of the
coupling. \par

As will be explained in the following, a crucial feature of this
procedure, which corresponds to momentum subtraction renormalisation
in continuum QCD, is that renormalised Green's functions do not depend
on the way the theory is regularised. As a consequence, LPTH is not
needed in order to relate the measured coupling to $\asms$, and the
relation between the two schemes can be computed entirely in continuum
perturbation theory. An analogous non-perturbative approach has been
recently applied, with encouraging results, to the renormalisation of
composite fermion operators~\cite{noi}.

The aim of this paper is to demonstrate the feasibility of our 
approach and to investigate the r\^{o}le of systematic
lattice uncertainties, such as discretisation effects and volume
dependence.  We perform this investigation in the quenched
approximation at $\beta = 6.0$ for two different lattice volumes.
Given the simplicity of the method, we expect the application to full
QCD to present no additional problems~\cite{ukqcd_in_progress}.

The paper is organised as follows: in Section \ref{section:method} we
present
the renormalisation scheme, the definition of the coupling and the numerical 
procedure.
In Section \protect\ref{section:lattice} we 
give the lattice results, discussing all sources of systematic errors in the calculation. In Section \ref{section:matching} we discuss
the procedure to relate our coupling to $\asms$ 
and we compare our non-perturbative results with predictions based on the use of LPTH. In Section \ref{section:conclusions} we summarise our work and we discuss some  possible future developments. 
Finally, the details of the perturbative
calculations 
are given in the appendix.

\section{The Method}
\label{section:method}
Provided systematic errors are under control, in order to compute $\asms$
from
the lattice it is sufficient to: 
\begin{enumerate}
\item set the scale of momenta in physical units 
by determining the lattice spacing $a$; 
\item define a suitable renormalisation scheme and a renormalised coupling to 
be measured;
\item match the result to the $\MSB~$ scheme.
\end{enumerate}
The physical quantities most frequently used
to determine the value of $a$ are: the $\rho$ meson mass, the string
tension, the 1P-1S mass splitting in heavy quarkonia~\cite{Aida}, 
and a characteristic length $r_0$, 
phenomenologically
connected to the intermediate range of the heavy quark potential~\cite{Sommer}.
Each choice has its theoretical and technical advantages,
extensively discussed in the literature.

In this work we set $\beta=6.0$ and we take the value of $a^{-1}$  
determined by Bali and Schilling~\cite{Bali} in their string tension
measurements. 
These yield $a^{-1} = 1.9 \pm 0.1 \ \Gev$.
We quote a systematic error on the scale to take into account the
uncertainty
resulting from other possible choices.

The definition and measurement of the coupling is achieved by
computing the gluon propagator and the three-gluon vertex function in a 
fixed gauge and 
renormalising them in a non-perturbative way. We define the lattice gluon 
field $A_{\mu}(x)$ as
\begin{equation} 
A_\mu(x+\hat{\mu}/2) = \frac{U_\mu(x) - U^\dagger_{\mu}(x)}{2i a g_0} 
- \frac{1}{3} {\rm Tr} \left( \frac{U_{\mu}(x) - 
U^\dagger_{\mu}(x)}{2i a g_0} \right), 
\label{eq:symmetric} 
\end{equation} 
where $\hat{\mu}$ indicates a unit lattice vector in the $\mu$ direction 
and $g_0$ is the bare coupling constant (we omit the colour index).

After performing the Fourier transform of (\ref{eq:symmetric}) one
can define (unrenormalised) lattice n-point gluon Green's functions, in momentum
space: 
\begin{equation}
G^{(n)}_{U \ \mu_1 \mu_2 \ldots \mu_n} (p_1, p_2, \ldots, p_{n})
= \langle A_{\mu_1}(p_1)A_{\mu_2}(p_2) \ldots A_{\mu_n}(p_n) \rangle,
\label{eq:npt}
\end{equation}
where $\langle \cdot \rangle$ indicates the Monte-Carlo average and momentum
conservation implies $p_1+p_2+\ldots +p_{n} = 0$.
 
Since the lattice calculation aims to evaluate such
Green's functions in the ``continuum window'', i.e. for a range of
parameters such that continuum physics is observed, the following 
criteria must be satisfied:
\begin{enumerate}
\item $\beta \geq 6.0 $, so that scaling is
observed for physical quantities.
\item $L a $ is large enough in physical units, where $L$ is the
linear dimension of the lattice, such that finite-volume effects are
under control.
\item Discretisation errors due to 
the contribution of nonleading terms in $a^2$ to the Green's
functions are negligible. 
\end{enumerate}
Since we work in momentum space, the last
requirement can be directly translated into a restriction on the range
of lattice momenta that should be used.
In the remainder of this section we assume that all the above requirements
are met and we adopt the formalism of continuum QCD.

We choose to work in the Landau gauge. 
As already mentioned, the quantities of interest 
are the 
unrenormalised gluon 
propagator
\begin{equation}
G^{(2)}_{U \ \mu \nu}(p) \equiv T_{\mu \nu}(p)\ G_{U} (p^2), \qquad 
T_{\mu \nu}(p) \equiv \delta_{\mu \nu} - \frac{p_{\mu} p_{\nu}}{p^2}
\label{eq:self}
\end{equation}
 and the complete unrenormalised 
gluon three-point function
$G_{U \ \alpha \beta \gamma}^{(3)} (p_{1}, p_{2}, p_{3})$.
This is related to the one-particle-irreducible three-point function
$\Gamma_{U \ \alpha \beta \gamma}^{(3)} (p_{1}, p_{2}, p_{3})$ by the
formula
\begin{equation}
G_{U \ \alpha \beta \gamma}^{(3)} (p_{1}, p_{2}, p_{3}) \equiv
\Gamma_{U \ \delta \rho \xi}^{(3)} (p_{1}, p_{2}, p_{3})
\  G_{U \ \delta \alpha }^{(2)} (p_1) \ G_{U \ \rho \beta }^{(2)} (p_2) 
\ G_{U \ \xi \gamma }^{(2)} (p_3).
\label{eq:ampu}
\end{equation}

In order to renormalise the gluon wave function, we impose that at a fixed
momentum scale $p^2=\mu^2$  the renormalised gluon propagator takes its
continuum tree-level value. 
One gets the non-perturbative renormalisation condition
\begin{equation}
G_R (p)\vert_{p^2=\mu^2}
= Z^{-1}_{A} (\mu a) G_U (p a)\vert_{p^2=\mu^2}
\nonumber \\
=\frac{1}{\mu^2}.
\label{eq:za}
\end{equation}
The above equation defines in a non-perturbative way the gluon
wave-function renormalisation $Z_A$.

For the three-gluon vertex, we start by choosing a kinematics
suitable for the lattice
geometry and which allows a simple non-perturbative 
definition of the vertex renormalisation constant $Z_V$.
Recalling the general form of $\Gamma^{(3)}_U$ in the continuum~\cite{general},
it turns out that if one evaluates $G^{(3)}_U$
in the Landau gauge at the kinematical points defined by
\begin{equation}
\alpha = \gamma,
\qquad p_{1}=-p_{3}=p, \ p_{2}=0,
\label{eq:asy}
\end{equation}
 then one can write 
\begin{equation}
\frac{\sum_{\alpha=1}^{4} \ G_{U \ \alpha \beta \alpha}^{(3)}
 (p a, 0, -p a)}{(G_{U} (p a))^2 \ G_{U} (0)} =
6 \ i \ Z_{V}^{-1} (pa) \ g_{0} \ p_{\beta}.
\label{eq:baba}
\end{equation}
With the above definition 
$Z_V$ contains a term which is linear in the external momenta but not proportional to the tree-level vertex 
(see appendix). Notice that the numerical calculation on the lattice 
for the three-gluon 
vertex yields directly the product $Z_V^{-1} g_0$. 

At this point we can define the running coupling $g$ at the scale $\mu$
from the renormalised three-gluon vertex at the asymmetric point as
\beq g(\mu) = Z_A^{3/2}(\mu a) \ Z_{V}^{-1} (\mu a) \ g_{0},
\protect\label{eq:renvert}
\eeq 
where the relevant renormalisation constants have been defined in
(\ref{eq:za}), (\ref{eq:baba}) and $\alpha_s(\mu) \equiv
g(\mu)^2/4 \pi$. 
This choice corresponds to a 
momentum subtraction scheme, usually referred 
to as $\MOM$ in continuum QCD~\cite{HH}.
 We postpone the discussion of the matching
procedure until Section~\ref{section:matching}.

\subsection{Computational Procedure}

$SU(3)$ gauge configurations were generated at $\beta = 6.0$ at two
lattice sizes; 150 configurations on a $16^4$ lattice, and $103$
configurations on a $24^4$ lattice.  
The configurations on the smaller lattice were generated on a 16K
CM-200 at the University of Edinburgh, using a hybrid-overrelaxed
algorithm, where both Cabibbo-Marinari pseudo-heatbath and overrelaxed
updates were performed on three $SU(2)$ subgroups.  Successive
configurations were separated by 150 sweeps (every sixth sweep a
heat-bath), 1000 sweeps being allowed for thermalisation.  Landau
gauge-fixing was achieved by using a Fourier-accelerated
algorithm~\cite{Davies}.  Autocorrelations were investigated by
performing a standard jackknife error analysis.

The data for the larger lattice were generated on the ACPMAPS 
supercomputer at FNAL using
the Creutz pseudo-heatbath algorithm, with 1600 sweeps between
configurations.  The configurations were fixed to the Landau gauge
using an overrelaxation algorithm, with the final iterations being
performed in double precision.  

A crucial step in the method is the accurate implementation of the
lattice Landau gauge condition
\begin{equation}
\Delta(x) = \sum_\mu A_\mu(x+\hat{\mu}) - A_\mu(x) = 0.
\protect\label{eq:lala}
\end{equation}
To monitor the gauge-fixing accuracy we compute the quantity
\begin{equation}
\theta = \frac{1}{V N_C} \sum_x {\rm Tr}\ \Delta^{\dag}(x) \Delta(x)
\end{equation}
as the algorithm progresses, terminating when $\theta < 10^{-11}$
(this is close to 32-bit machine precision). Since our calculations
involve low momentum modes of gluon correlation functions, the above
test is not in itself sufficient because $\Delta(x)$ is a local
quantity. For this reason, we also compute
\begin{equation}
A_0(t) = \sum_{\vec{x}} A_0(\vec{x}, t).
\end{equation}
In a periodic box, the Landau gauge condition implies that $A_0(t)$ is independent of $t$~\cite{Mandula}. 
In particular, for our configurations on the smaller lattice, 
$A_0(t)$ was constant to better than one part in $10^5$.
For the purpose of our analysis, the only quantity that needs to be
stored is the Fourier-transformed field $A_\mu(p)$ for a selected
range of lattice momenta. All $n$-point gluon correlation functions
can then be assembled using eq.~(\ref{eq:npt}), where momentum
conservation is imposed explicitly.

To compute $\alpha_s$, we first evaluate the gluon propagator and
determine $Z_A$ from eq.~(\ref{eq:za}). Next, we measure the complete
three-point function $G_{U}^{(3)}$ of the gluon field and the quantity
on the l.h.s.\ of (\ref{eq:baba}). Finally, $g(\mu)$ is obtained
from eq.~(\ref{eq:renvert}). We take advantage of all the symmetries
of the problem to improve statistics.  The quoted errors are obtained
using a single-point-elimination jackknife algorithm.

\section{Results}
\protect\label{section:lattice}
\subsection{Tensor Structure}

We start by analysing the tensor structure of the
lattice gluon propagator and three-gluon vertex function 
as a means both of determining the degree of violation 
of continuum rotational invariance, and of verifying the extent
to which the Landau gauge condition is satisfied in momentum space.

It is worth noting that our definition (\ref{eq:symmetric}) for the gluon 
field differs from the one which has been used in all non-perturbative 
calculations to date, which is~\cite{Mandula}: 
\begin{equation}
A'_{\mu}(x) = \frac{U_{\mu}(x) - U^\dagger_{\mu}(x)}{2i a g_0} - \frac{1}{3}
{\rm Tr} \left( \frac{U_{\mu}(x) - U^\dagger_{\mu}(x)}{2i a g_0} \right).
\protect\label{eq:defi}
\end{equation}
It turns out that 
the above ``asymmetric'' definition is not consistent with the one 
which is usually used in perturbative lattice calculations.
Consistency is achieved by using the ``symmetric'' definition 
(\ref{eq:symmetric}), which has better properties in the continuum limit. 
To illustrate this point, we observe that in momentum space the two 
definitions are related by the formula
\begin{equation}
A_\mu(p) = e^{-i p_\mu/2} A^{'}_\mu(p).
\protect\label{eq:symmetric_p}
\end{equation}
If we now write 
the Landau gauge-fixing condition (\protect\ref{eq:lala}) in momentum space, 
using the  
asymmetric definition one gets
\begin{equation}
\sum_\mu \left[ (\cos p_\mu - 1) - i \sin p_\mu \right]
A^{'}_\mu(p)= 0,
 \label{eqn:asym_landau_p}
\end{equation}
while the symmetric definition yields
\begin{equation}
\sum_\mu 2 i \sin p_\mu/2 \, A_\mu(p) = 0. 
\label{eqn:sym_landau_p}
\end{equation}
In the limit $a \rightarrow 0$ the continuum gauge 
condition is recovered with $O(a)$ corrections in the asymmetric case and 
$O(a^2)$ in the symmetric one. Thus the latter corresponds to 
an ``improved'' lattice Landau gauge condition.

By using the symmetric definition one can check very 
accurately the tensor structure of non-perturbative lattice Green's functions 
against what is expected from LPTH. 
Based on such a definition  
we expect the Landau gauge propagator to satisfy
\begin{equation}
G^{(2)}_{U \ \mu \nu}(p) = \hat{T}_{\mu \nu}(\hat{p})\ G_{U} (\hat{p}^2),
\qquad \hat{T}_{\mu \nu}(\hat{p}) \equiv \delta_{\mu \nu} -
\frac{\hat{p}_{\mu} \hat{p}_{\nu}}{\hat{p}^2}, 
\label{eq:proplat}
\end{equation}
where
\begin{equation}
\hat{p}_{\mu} = \frac{2}{a} \sin (\frac{p_{\mu} a}{2}).
\end{equation}
 This makes  
the analysis of violations of rotational invariance 
very simple since, if our propagator is found to satisfy
(\ref{eq:proplat}), one has for any $p \neq0$: 
\begin{equation}
\sum_{\mu} \  G^{(2)}_{U \ \mu \mu}(p) = 3 \ G_{U} (\hat{p}^2),
\label{eq:tra}
\end{equation}
so that violations of rotational invariance in $\sum_{\mu} \  
G^{(2)}_{U \ \mu \mu}(p)$  
can only arise from the scalar part $G_U(p^2)$.
If the asymmetric definition is used, the simple tensor structure in
eq.~(\ref{eq:proplat}) is modified and the interpretation of the numerical 
results is more difficult.  

In Tables
\protect\ref{tab:o4_sym_16^4} and \protect\ref{tab:o4_asym_16^4} we
show ratios of tensor components of the gluon propagator on the $16^4$
lattices using the symmetric and asymmetric definitions of $A_{\mu}$
respectively. These are compared to what is expected from
(\ref{eq:proplat}) and its continuum counterpart.  In all cases the
uncertainty is less than one unit in the last quoted figure.
\begin{table}[t]
\begin{center}
\begin{tabular}{|c|ccc|ccc|ccc|}\hline
$p$ & $\frac{p_0}{p_1}$ & $\frac{\sin p_0/2}{\sin p_1/2}$ & 
$-\frac{G_{01}}{G_{00}}$ & $\frac{p_1}{p_0}$ & 
$\frac{\sin p_1/2}{\sin p_0/2}$ & $-\frac{G_{01}}{G_{11}}$ &
$\frac{p_1^2}{p_0^2}$ &
$\frac{\sin^2 p_1/2}{\sin^2 p_0/2}$ & $\frac{G_{00}}{G_{11}}$ \\[0.5ex] \hline
$(1,1,0,0)$ & 1 & 1 & 1.00 & 1 & 1.0 & 1.00 & 1 & 1.0 & 1.00 \\
$(1,2,0,0)$ & $1/2$ & 0.510 & 0.510 & 2 & 1.962 & 
1.962 & 4 & 3.849 & 3.848 \\
$(2,2,0,0)$ & 1 & 1 & 1.00 & 1 & 1.0 & 1.00 & 1 & 1.0 & 1.00 \\
$(1,3,0,0)$ & $1/3$ & 0.351 & 0.351 & 3 & 2.848 & 2.848 & 9 &
8.110 & 8.110 \\ 
$(3,3,0,0)$ & 1 & 1 & 1.00 & 1 & 1.0 & 1.00 & 1 & 1.0 & 1.00 \\ \hline
\end{tabular}\\[1.0ex]
\caption{Symmetry tests for $G^{(2)}_{U \ \mu \nu}(p)$ 
on the $16^4$ lattices at $\beta = 6.0$, using the symmetric
definition of $A_\mu$.}
\protect\label{tab:o4_sym_16^4}
\end{center}
\end{table}
\begin{table}[t]
\begin{center}
\begin{tabular}{|c|ccc|ccc|ccc|}\hline
$p$ & $\frac{p_0}{p_1}$ & $\frac{\sin p_0/2}{\sin p_1/2}$ & 
$-\frac{G_{01}}{G_{00}}$ & $\frac{p_1}{p_0}$ & 
$\frac{\sin p_1/2}{\sin p_0/2}$ & $-\frac{G_{01}}{G_{11}}$ &
$\frac{p_1^2}{p_0^2}$ &
$\frac{\sin^2 p_1/2}{\sin^2 p_0/2}$ & $\frac{G_{00}}{G_{11}}$ \\[0.5ex] \hline
$(1,1,0,0)$ & 1 & 1 & 1.00 & 1 & 1 & 1.00 & 1 & 1 & 1.00 \\
$(1,2,0,0)$ & $1/2$ & 0.510 & 0.50 & 2 & 1.962 & 1.924 & 4 & 3.848 &
3.848 \\
$(2,2,0,0)$ & 1 & 1 & 1.00 & 1 & 1 & 1.00 & 1 & 1 & 1.00 \\
$(1,3,0,0)$ & $1/3$ & 0.351 & 0.324 & 3 & 2.848 & 2.631 & 9 & 8.110 &
8.110 \\ 
$(3,3,0,0)$ & 1 & 1 & 1.00 & 1 & 1 & 1.00 & 1 & 1 & 1.00 \\ \hline
\end{tabular}
\caption{
Symmetry tests for $G^{(2)}_{U \ \mu \nu}(p)$ 
on the $16^4$ lattices at $\beta = 6.0$, using the asymmetric
definition of $A_\mu$.
}
\protect\label{tab:o4_asym_16^4}
\end{center}
\end{table}
We note the following:
\begin{itemize}
\item For the symmetric definition of $A_{\mu}$, the numerical data 
are completely consistent with our expectation from eq.~(\ref{eq:proplat}). 
This indicates that 
the lattice gauge condition has been implemented very accurately, so that  
we have a complete
understanding of the tensor structure of the two-point
function.
\item For the asymmetric definition of $A_{\mu}$, the picture is quite
different, although even in this case the deviation from
(\ref{eq:proplat}) is at most 10\% for the momenta shown.
\end{itemize}
\bigskip

We consider now the three-gluon correlator 
$G^{(3)}_{U \ \alpha\beta\gamma}(p,0,-p).$ 
On the lattice we expect it to satisfy
\begin{equation}
G^{(3)}_{U \ \alpha\beta\gamma}(p,0,-p) =
\hat{T}_{\alpha\gamma}(\hat{p}) \ F(\hat{p}^2) \ \hat{p}_{\beta}
 \ (G_{U} (\hat{p}^2))^2 \ G_{U} (0), 
\label{eqn:landau_threept}
\end{equation}
which is the form one obtains by 
performing the substitution $p_\mu \rightarrow \hat{p}_{\mu}$ in 
the continuum expression. Here $F(\hat{p}^2)$ represents a generic
function of $\hat{p}^2$.
Note that from (\protect\ref{eq:symmetric_p}) it follows that 
$G^{(3)}_{U \ \alpha\beta\gamma}(p,0,-p)$ does not depend 
on the definition used for the gauge field as the  
extra phase factor associated with the symmetric
definition of $A_\mu(p)$ cancels at this kinematic point.

In Tables \protect\ref{tab:o4_sym_3pt} and \protect\ref{tab:o4_sym_3pt_24} 
we show ratios of tensor components
 of $G^{(3)}_{U \ \alpha\beta\gamma}(p,0,-p)$ on the $16^4$ and $24^4$ 
lattices
compared to expectations from (\ref{eqn:landau_threept})
and its continuum counterpart.
\begin{table}[htb]
\begin{center}
\begin{tabular}{|c|ccc|ccc|ccc|}\hline
$p$ & $\frac{ p_1^2}{p_0^2}$ & $\frac{\sin^2 p_1/2}{\sin^2 p_0/2}$ &
$\frac{G_{010}}{G_{111}}$ & $\frac{p_0}{p_1}$ & 
$\frac{\sin p_0/2}{\sin p_1/2}$ & $\frac{G_{101}}{G_{111}}$ & 
$\frac{p_1}{p_0}$ & $\frac{\sin p_1/2}{\sin p_0/2}$ &
$-\frac{G_{011}}{G_{111}}$ \\[0.5ex] \hline
$(1,1,0,0)$ & 1 & 1 & 1.000 & 1 & 1 & 1.1(2) & 1 & 1 & 1.000 \\
$(1,2,0,0)$ & 4 & 3.848 & 3.848 & $1/2$ & 0.510 & 0.3(1) & 2 & 1.962 &
1.962 \\
$(2,2,0,0)$ & 1 & 1 & 1.000 & 1 & 1 & 0.8(3) & 1 & 1 & 1.000 \\ \hline
\end{tabular}\\[1.0ex]
\caption{Symmetry tests for $G^{(3)}_{U \ \alpha\beta\gamma}(p,0,-p)$
 in the Landau gauge, using
the symmetric definition of $A_\mu$, on $16^4$ lattices.  Unless
otherwise noted, the error is always less than one unit in the last
quoted figure.}
\protect\label{tab:o4_sym_3pt}
\end{center}
\end{table}

\begin{table}[htb]
\begin{center}
\begin{tabular}{|c|ccc|ccc|ccc|}\hline
$p$ & $\frac{ p_1^2}{p_0^2}$ & $\frac{\sin^2 p_1/2}{\sin^2 p_0/2}$ &
$\frac{G_{010}}{G_{111}}$ & $\frac{p_0}{p_1}$ &
$\frac{\sin p_0/2}{\sin p_1/2}$ & $\frac{G_{101}}{G_{111}}$ &
$\frac{p_1}{p_0}$ & $\frac{\sin p_1/2}{\sin p_0/2}$ &
$-\frac{G_{011}}{G_{111}}$ \\[0.5ex] \hline
$(1,1,0,0)$ & 1 & 1 & 1.000 & 1 & 1 & -0.3(4) & 1 & 1 & 1.000 \\
$(1,2,0,0)$ & 4 & 3.932 & 3.932 & $1/2$ & 0.504 & -0.2(4) & 2 & 1.983 &
1.983 \\
$(2,2,0,0)$ & 1 & 1 & 1.000 & 1 & 1 & 0.3(7) & 1 & 1 & 1.000 \\
$(1,3,0,0)$ & 9 & 8.596 & 8.596 & $1/3$ & 0.341 & 0.8(7) & 3 & 2.932 &
2.932 \\\hline
\end{tabular}\\[1.0ex]
\caption{Symmetry tests for $G^{(3)}_{U \ \alpha\beta\gamma}(p,0,-p)$
in the Landau gauge, using
the symmetric definition of $A_\mu$, on $24^4$ lattices.  Unless
otherwise noted, the error is always less than one unit in the last
quoted figure.}
\protect\label{tab:o4_sym_3pt_24}
\end{center}
\end{table}
The theoretical expectation is satisfied very accurately for two of the three 
ratios under consideration, while  
 the agreement for the 
ratio $G_{101}/G_{111}$ is much poorer, especially on the larger lattice.
This is related to the fact that the first two ratios are completely 
determined by the
Landau gauge condition for our choice of the momenta, whilst the
last ratio is not. Thus these results provide further evidence that 
the momentum-space Landau gauge-fixing condition is very well 
satisfied on all our lattices, but they also hint at a poorer quality 
of the data on the larger lattice.
 
\subsection{Renormalisation Constants and Running Coupling}
We are now left with the task of computing the renormalisation
constants defined in (\ref{eq:za}) and (\ref{eq:baba}) and the running
coupling (\ref{eq:renvert}). 
Notice that because of our definitions and choice of kinematics,
the difference between the symmetric and asymmetric definitions for the gauge 
field is immaterial for this purpose.
All data
are plotted vs.\ $\mu = \sqrt{p^2}$, expressed in $\Gev$. 
In order to detect violations of
rotational invariance, we have used whenever possible different
combinations of lattice vectors for a fixed value of $p^2$ and we have
plotted separately the corresponding data points.

\begin{figure}
\vspace{-1cm}
\centerline{\ewxy{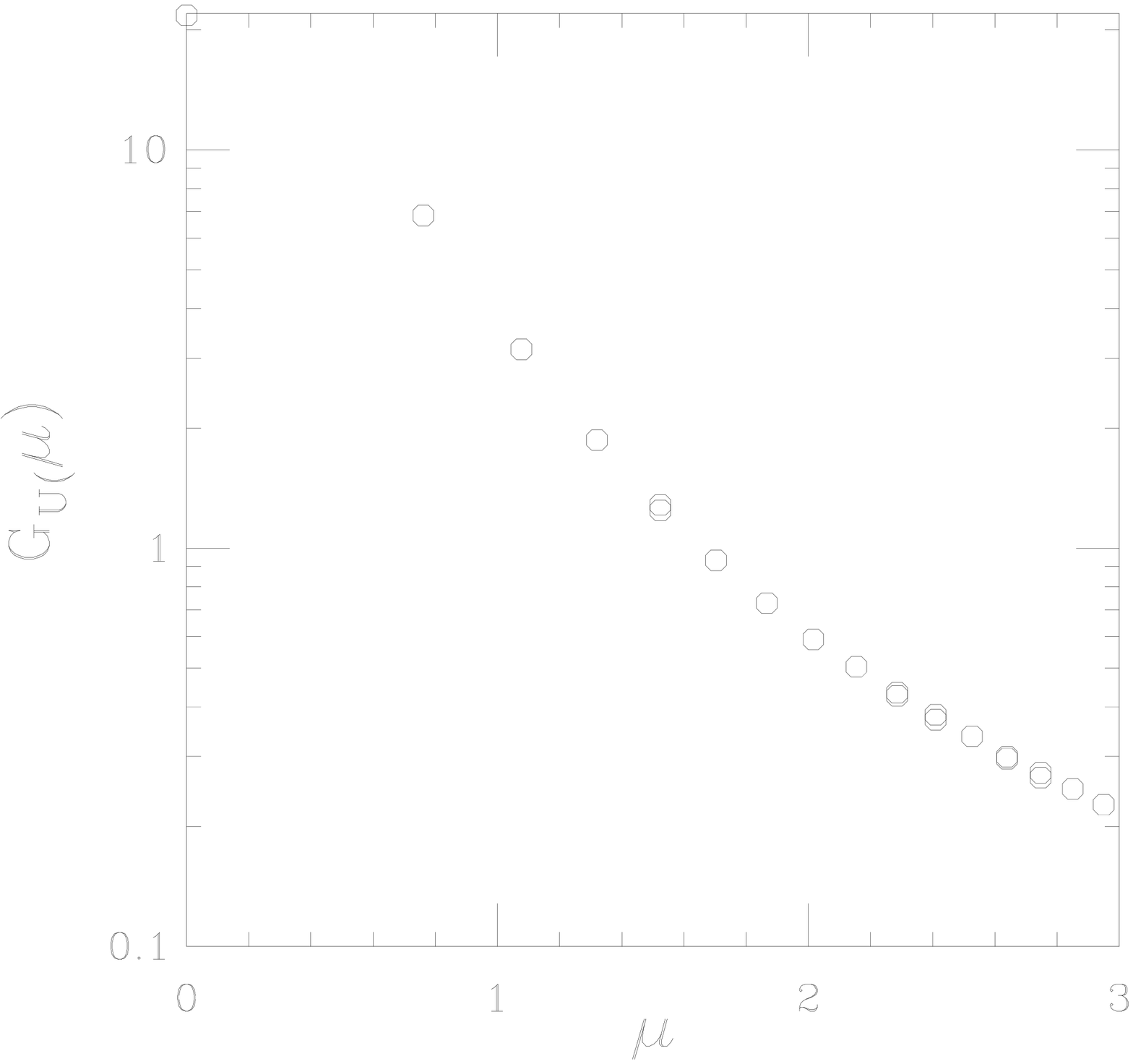}{256pt}}
\caption{$G_U(\mu)$ vs. $\mu$ 
for the $16^4$ lattices
at $\beta=6.0$.}
\protect\label{fig:glprop}
\end{figure}

In Figure~\ref{fig:glprop} we show the scalar gluon self-energy
$G_U(\mu)$.  Statistical errors are negligible and no
violations of rotational invariance can be detected in the momentum
range under consideration.
\begin{figure}
\epsfxsize=190pt \epsfbox{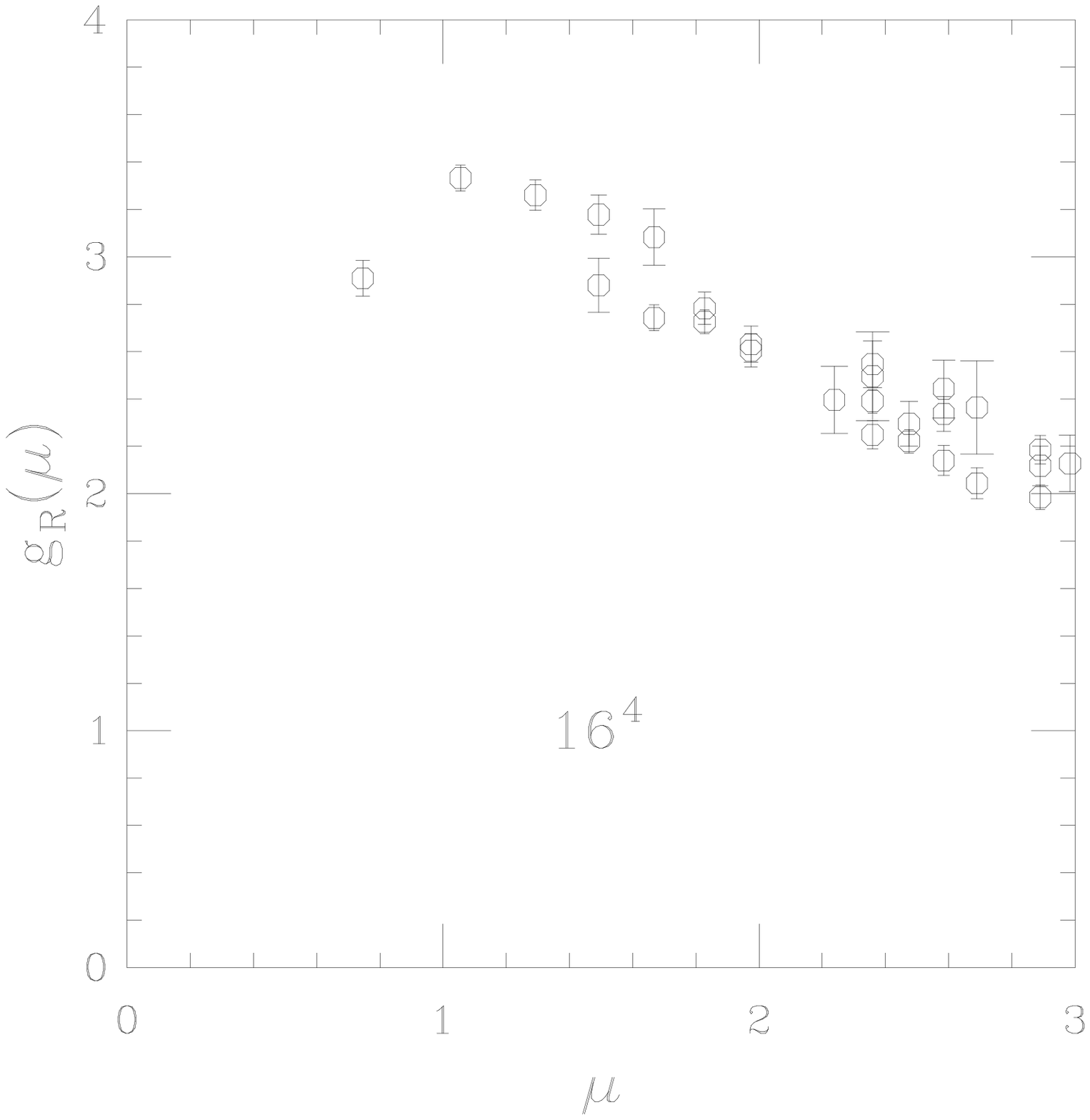} \hfill \epsfxsize=190pt
\epsfbox{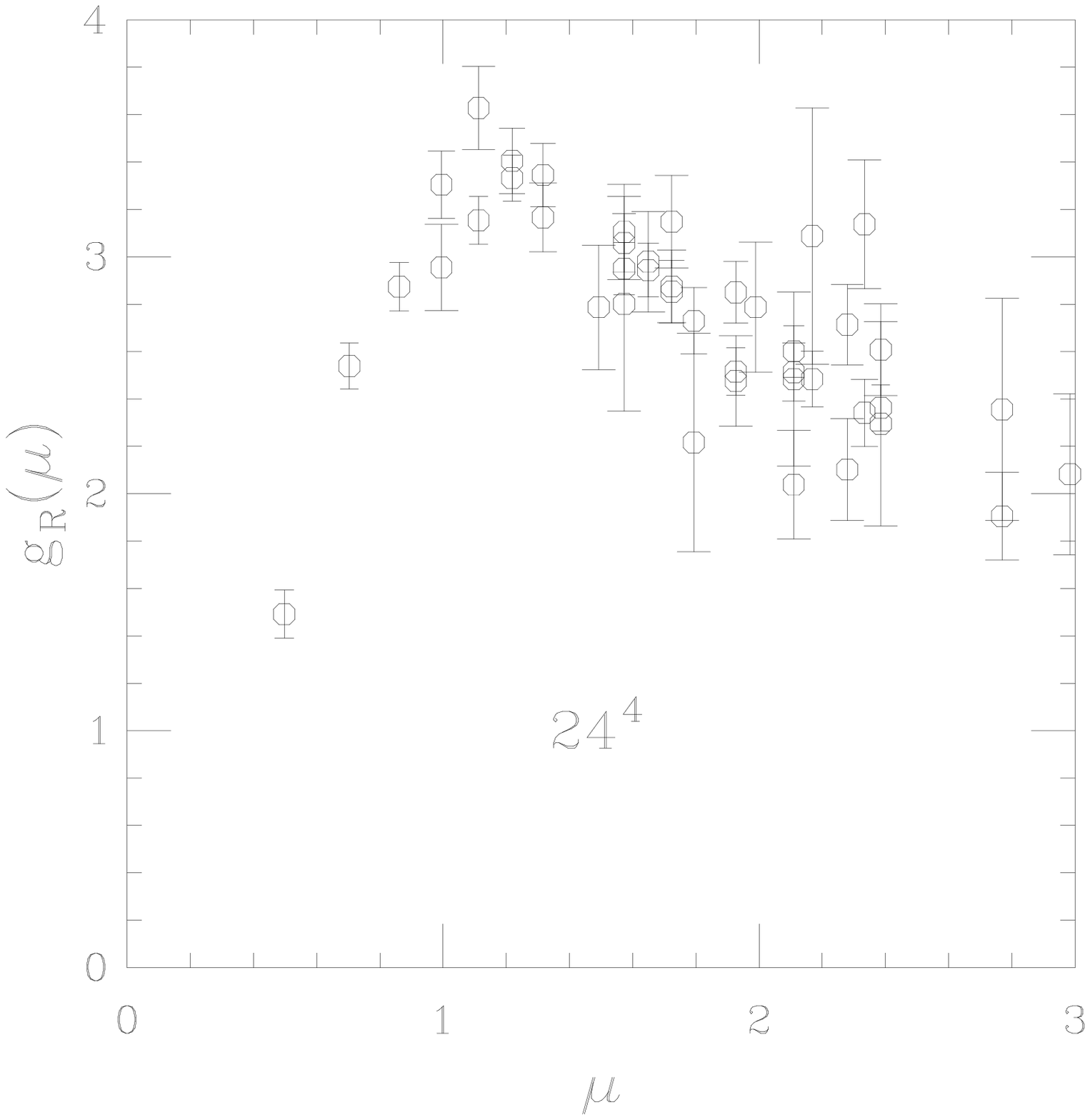} 
\caption{Running coupling $g(\mu)$ vs.\ $\mu$.}
\protect\label{fig:run}
\end{figure}
Next we compute the gluon three-point function and evaluate the
running coupling according to (\protect\ref{eq:renvert}).  This is
plotted in Figure~\protect\ref{fig:run} for both lattice sizes.  We
obtain again a clear signal; on the other hand, the data do show some
violations of rotational invariance. We note again that the errors on
the $24^4$ lattice are considerably larger than on the smaller
lattice.

\subsection{Systematic Lattice Uncertainties and Extraction of $\Lambda$}
The important physical question is whether one can isolate a range of
momenta where lattice artifacts are negligible and our coupling runs
according to the two-loop perturbative expression
\begin{equation}
g^2(\mu) = \left[b_0\ln(\mu^2/\Lam^2)+
\frac{b_1}{b_0}\ln\ln(\mu^2/\Lam^2) \right]^{-1},
\label{eq:twolooprun}
\end{equation}
where $b_0=11/16\pi^2$, $b_1=102/(16 \pi^2)^2$ and $\Lam$ is the QCD
scale parameter for the renormalisation scheme that we are using (in
the quenched approximation).  To answer this question, and obtain an
estimate for $\Lam$, we compute $\Lam$ as a function of the measured
values of $g^2(\mu)$ according to the formula
\begin{equation}
\Lam = \mu \ {\rm exp} \left(-\frac{1}{2 b_0 g^2(\mu)}\right)
\left[ b_0 g^2(\mu)\right]^{-\frac{b_1}{2 b_0^2}}.
\label{eq:twolooplam}
\end{equation}
If the coupling runs according to (\ref{eq:twolooprun}), then $\Lam$
as defined from the above equation must be constant.  Given the
exponential dependence of $\Lam$ on $g^2(\mu)$, this test is a very
stringent one.  As an alternative procedure one could fit the data for
$g^2(\mu)$ to formula (\ref{eq:twolooprun}), with $\Lam$ as a
parameter, but this would not result in a strong test since
(\ref{eq:twolooprun}) only depends logarithmically on $\Lam$.

By plotting $\Lam$ versus $\mu$ (see Figure~3), three different
regimes can be identified:
\begin{figure}
\vspace{-1cm}
\centerline{\ewxy{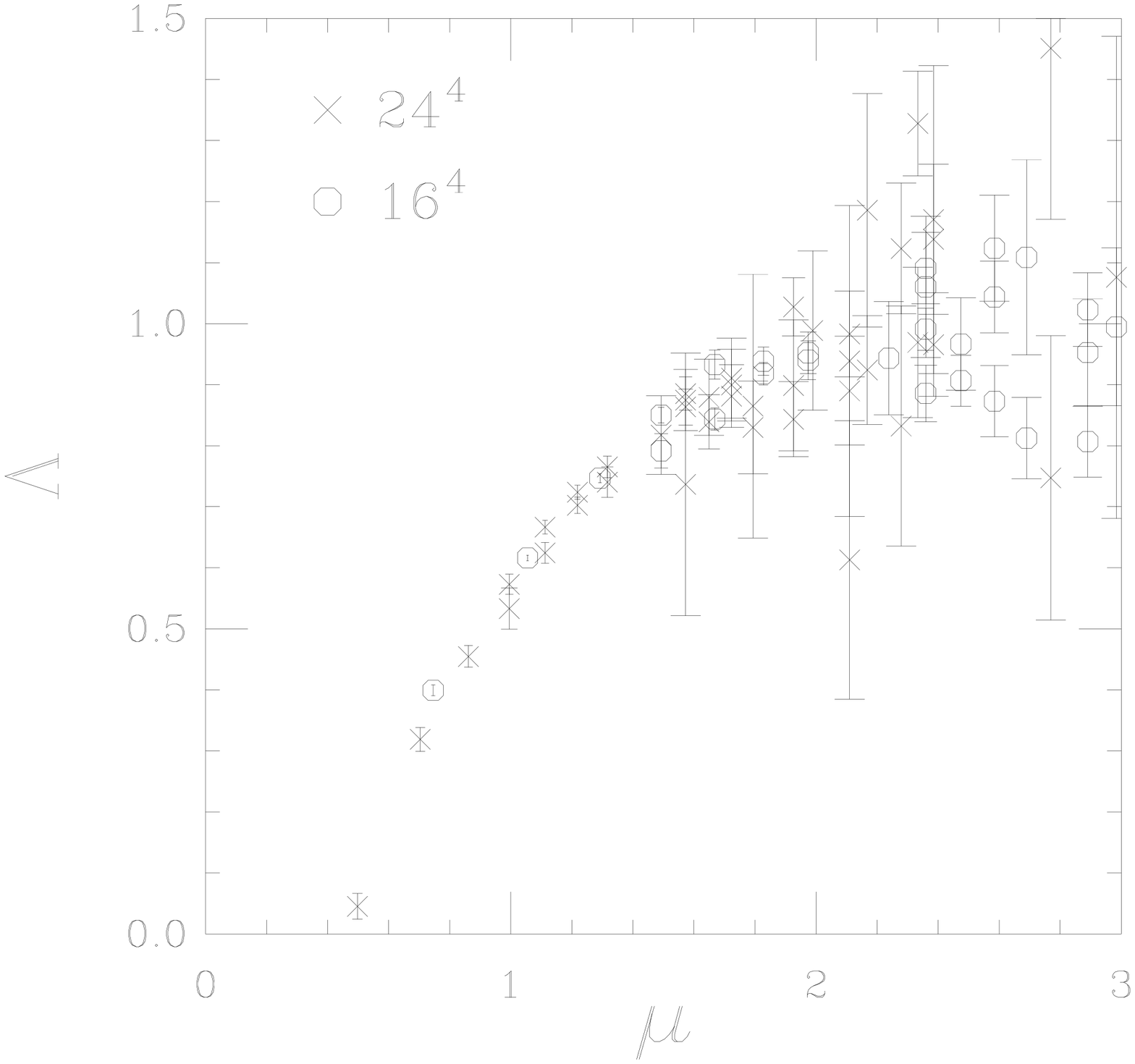}{256pt}}
\protect\label{fig:lambda}
\caption{$\Lam$ vs.\ $\mu$ for our two lattice sizes.}
\end{figure}
\begin{enumerate}
\item For $\mu < 1.8 \ \Gev$, $\Lam$ displays a strong dependence on the 
renormalisation scale. This is not surprising, as for low momenta 
asymptotic scaling is not expected.
\item In the range $1.8 < \mu < 2.3 \ \Gev$ the data are consistent with a 
constant value for $\Lam$. No violations of rotational invariance are 
observed in such a range and a comparison of the two lattice sizes shows 
no volume dependence either. This is shown in Figure~\protect\ref{fig:win}.
\item For $\mu > 2.3 \ \Gev$, rotational invariance is broken by higher 
order terms in $a^2$ and the two-loop behaviour disappears.
\end{enumerate}
\begin{figure}
\epsfxsize=190pt \epsfbox{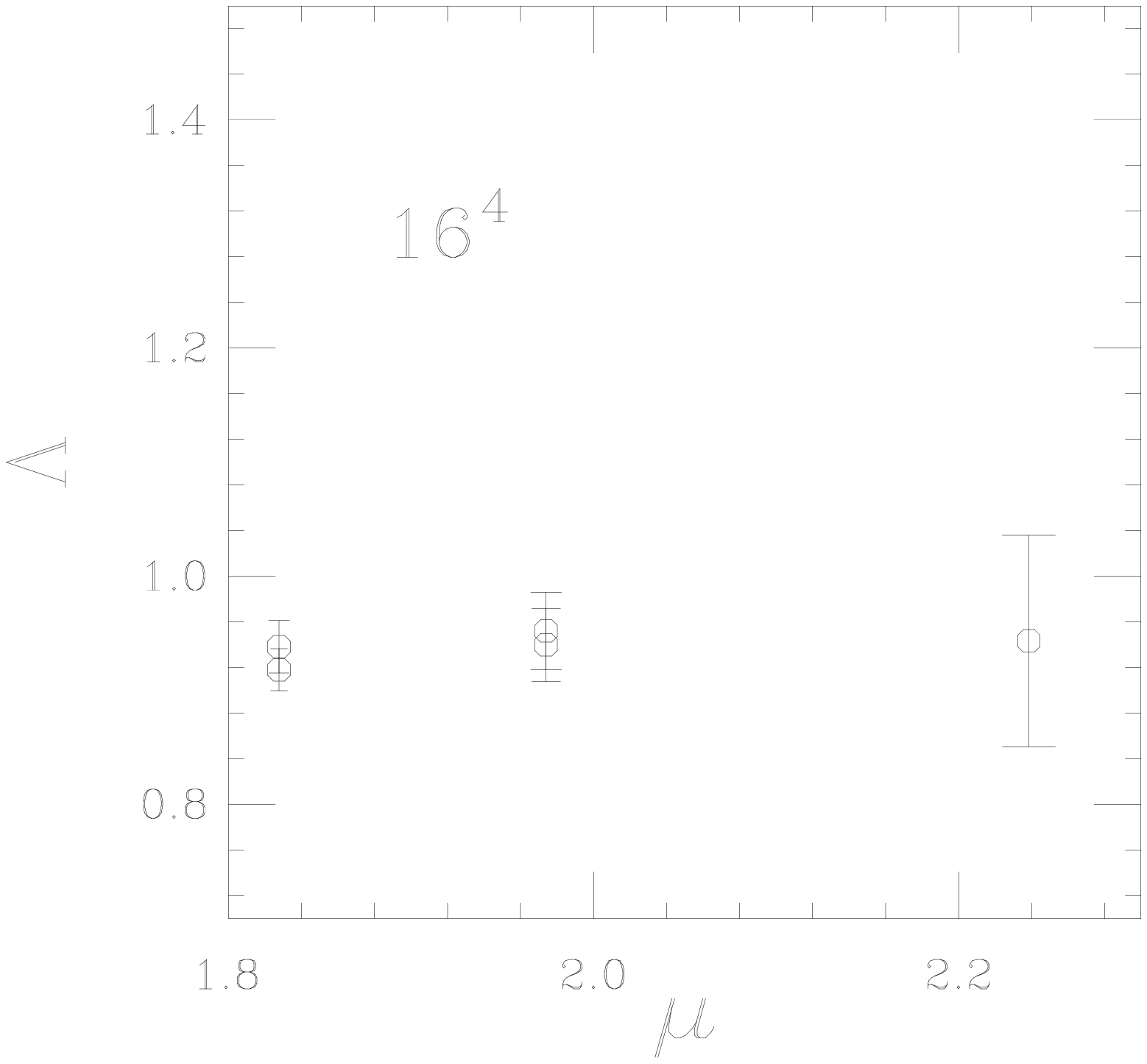} \hfill \epsfxsize=190pt
\epsfbox{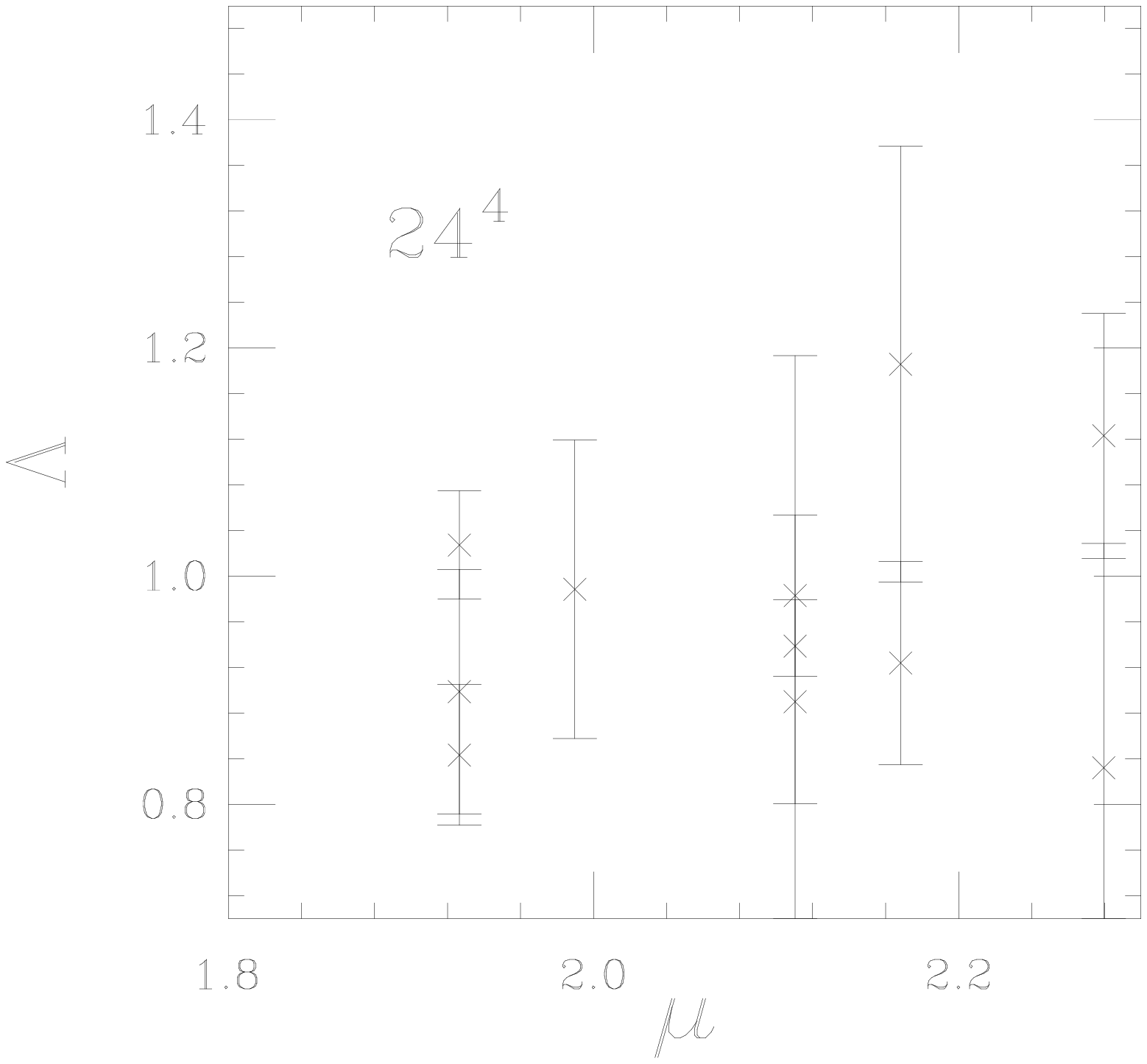}
\caption{$\Lam$ vs.\ $\mu$ in the ``continuum window''.}
\protect\label{fig:win}
\end{figure}

In summary, we appear to have a ``continuum window'' in the range 
$1.8 < \mu < 2.3  \ \Gev$, where two-loop
scaling is observed and lattice artifacts are under control.
In order to extract a prediction for $\Lam$, we fit the data points in
the continuum window to a constant.  We take as our best estimate
the fit to the $16^4$ data, for which the statistical errors are
smaller, and obtain
\beq 
\Lam=0.96 \pm 0.02
\pm 0.09 \ \Gev,
\label{eq:resu}
\eeq
where the first error is statistical and the second error comes from the
uncertainty on the value of $a^{-1}$. 

Since our lattice calculation is performed in a fixed gauge, 
it is worth mentioning that non-perturbative gauge-fixing ambiguities 
(Gribov copies) may in principle affect our results. These could be 
regarded as a potential source of systematic errors. 
However, based on previous investigations of other lattice gauge-fixed 
quantities\cite{copies}, 
we believe that the contribution of ``Gribov noise'' to the overall 
error, if at all present, is negligible. 

\section{Matching to $\overline{MS}$}
\label{section:matching}
\par
In this section we extract a prediction for $\asms$ with zero active quark 
flavours from our numerical 
results for $\Lam$.
As already mentioned, the procedure that we adopt avoids the use of
LPTH. We also discuss a LPTH calculation which provides some important
consistency checks.

We start from our numerical estimate (\eq{eq:resu}) of the continuum
scale parameter $\Lam$. 
The general relation between the scale parameters 
in two continuum schemes $A$ and $B$ can be written as
\beq 
\frac{\Lambda_A}{\Lambda_B}=\exp \Bigl[
- \frac{1}{2 b_0} \Bigl( \frac{1}{g_A^2(\mu)}
-\frac{1}{g_B^2(\mu)} \Bigr) + O(g^2(\mu)) \Bigr],
\label{eq:ratio}
\eeq 
with 
\beq g_A^2(\mu)=g_B^2(\mu)\Bigl( 1 + \frac{g_B^2(\mu)}{4
\pi} C_{AB} + O(g_B^2(\mu) \Bigr),
 \protect\label{eq:gengen} 
\eeq 
where
$C_{AB}$, which in general depends on the chosen gauge and on the
number of active flavours, is obtained from a continuum perturbative
calculation as described in the appendix.
Since $\Lambda_{A}, \ \Lambda_{B}$ are independent of $\mu$ and because of 
asymptotic freedom, by letting $\mu \rightarrow 
\infty$
 the ratio (\ref{eq:ratio}) can be determined to all orders in the coupling 
constant from the one-
loop calculation (see for example~\cite{Cel,Bil}). 

We obtain, in the Landau gauge and for zero quark flavours
\beq
\frac{\Lams^{(0)}}{\Lam^{(0)}}=0.35.
\label{eq:matchcont}
\eeq
This result  is in agreement with 
with previous one-loop calculations of the three-gluon vertex~\cite{BF}.

Note that the scheme that we have adopted for the non-perturbative
calculation (see Section 2) 
differs from the usual $\MOM$ scheme as it contains an
extra constant term in the
vertex renormalisation constant.
This is a linear term in the momenta, not
proportional to the tree level vertex, which is equal on the lattice and in
the
continuum. The coefficient in (\ref{eq:matchcont})
includes it perturbatively (see appendix).
\par
Using (\eq{eq:resu}) and (\ref{eq:matchcont}), we can
extract $\Lams^{(0)}$ from
\beq
\Lams^{(0)}=\Lam^{(0)} \ \frac{\Lams^{(0)}}{\Lam^{(0)}}.
\eeq
We get
\beq
\Lams^{(0)}= 0.34 \pm 0.05 ~\Gev.
\eeq
This is the main result of our computation.

Our result can be directly compared with the one of ref.~\cite{Bali}:
\beq 
\Lams^{(0)}= 0.293 \pm 0.018 ^{+0.025}_{-0.063} ~\Gev.  
\eeq
In terms of $\asms^{(0)}$, our result yields:
\begin{equation}
\asms^{(0)}(2.0 ~\Gev)= 0.25 \pm 0.02.
\end{equation}
We do not attempt to estimate $\asms^{(n_f)}$ for $n_f \ne 0$ on the
basis of quenched data; a computation in full QCD with two degenerate
flavours of sea quarks is in progress.

\subsection{Starting from the Lattice}
\label{section:latt_matching}
One can apply renormalisation group considerations, analogous to the ones
that establish the momentum dependence of the renormalised coupling,
to the bare theory.
On the lattice, as in any other regularisation scheme, 
renormalisability implies that the bare coupling constant should be
cut-off dependent and, for $a$ small enough, $g_0=g_0(a)$ should be a
universal function of $a$.
Thus in the scaling region one can define,
up to an arbitrary integration constant,
an $a$-independent $\Lamlatt$ parameter,
in terms of the lattice spacing and of the bare lattice coupling
as
\beq
\Lamlatt=\frac{1}{a} \exp \Bigl(\int^{g_0(a)}
\frac{dg'_0} {\beta(g'_0)}\Bigr).
\label{eq:scaling}
\eeq
In the asymptotic scaling region $a 
\rightarrow 0$ and $g_0(a) \rightarrow  0$, where the $\beta$-function
is perturbatively computable, one can fix the integration constant by
defining $\Lamlatt$ as
\beq
\Lamlatt=\frac{1}{a} \exp \Bigl( - \frac{1}{2 b_0 g^2_0(a)} \Bigr)
\left( b_0 g^2_0(a) \right)^{- \frac{b_1}{2 b_0^2}},
\label{eq:lamlatt}
\eeq independent of $a$ in the asymptotic scaling limit.  Note that,
whereas we find asymptotic scaling for the renormalised coupling for
momenta $1.8 < q < 2.3 \ \Gev$, 
see eq.~(\eq{eq:twolooplam}) and Figure~3, this
does not necessarily imply that the asymptotic regime has already set
in for the bare coupling used in the simulation.  As a check, which
involves lattice perturbation theory, we can \begin{enumerate}
\item take the non-perturbative determination of $\Lam$,
\item compute the ratio
$\Lamlatt / \Lam$ 
in lattice perturbation theory 
\item extract $\Lamlatt$ from
\beq
\Lamlatt=\Lam \ \frac{\Lamlatt}{\Lam}
\label{eq:matchlatt}
\eeq
\end{enumerate}
and compare the value obtained from eq.~(\eq{eq:matchlatt}) with
the one from eq.~(\eq{eq:lamlatt}).
\par
The ratio
\beq
\frac{\Lam}{\Lamlatt}= \mu a\exp \Bigl[
- \frac{1}{2 b_0} \Bigl( \frac{1}{\gdue(\mu)}
-\frac{1}{g_0^2(a)} \Bigr) + O(g_0^2) \Bigr]
\label{eq:ratio2}
\eeq
has been calculated at one-loop by Hasenfratz and
Hasenfratz~\cite{HH} in the Feynman gauge $\lambda=1$, where
$\lambda$ here is the gauge parameter. They found
\beq
\frac{\Lam(\lambda=1)}{\Lamlatt}= 69.4.
\eeq
We have checked their result by working in a general covariant gauge.
The details of the LPTH calculation are given in the appendix.
In the Landau gauge, we get
\beq
\frac{\Lam(\lambda=0)}{\Lamlatt}= 54.6.
\eeq
The number that we need to insert in (\ref{eq:matchlatt}) is not quite the
above one, because of the already mentioned
extra term in the vertex renormalisation,
not proportional to the tree level vertex.
In our scheme the result is
\beq
\frac{\Lam(\lambda=0)}{\Lamlatt}= 83.2.
\label{eq:lpres}
\eeq
By inserting it into eq.~(\eq{eq:matchlatt}), and using the result
(\eq{eq:resu}), one would get
\beq 
\Lamlatt = 11.6 ~\Mev,
\eeq
to be compared with
\beq
\Lamlatt = 4.5 ~\Mev
\eeq
obtained from the hypothesis of asymptotic scaling, eq.~(\eq{eq:lamlatt}),
with $g_0^2=1$ and $a^{-1}=1.9 ~\Gev$.
The comparison 
confirms the well known result that for values of 
$\beta$
accessible to current simulations, $\Lamlatt$ still displays 
$\beta$ dependence~\cite{Bali}.
We stress again that in our case the matching
procedure, described in the previous subsection,
does not require knowledge of $\Lamlatt$.
\par
Another way of seeing the failure of LPTH in this
case is to compare the non-perturbative results for $g^2(\mu)$ 
with what is obtained by inserting in the relation
\beq
g^2(\mu)=Z_A^{3}(\mu a) Z_V^{-2}(\mu a)
g_0^2(a) 
\eeq
the values of the $Z$'s obtained in LPTH
and ``boosted'' lattice perturbation theory~\cite{boost}. 
This comparison is shown in Figure~5.

\begin{figure}[t]
\vspace{2.4in}
\includegraphics{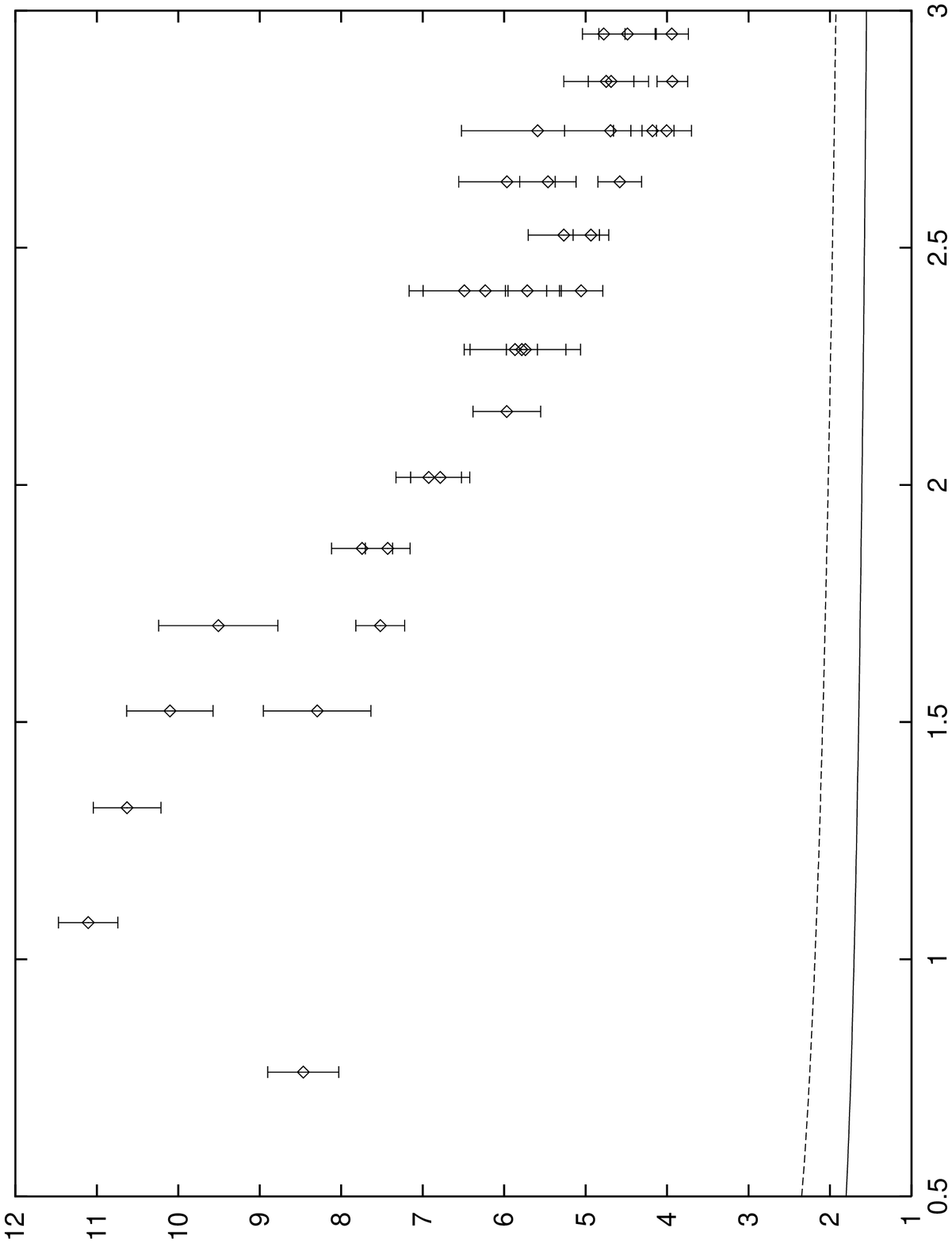}
\caption{$g^2(\mu)$ vs.\ momentum as evaluated non-perturbatively on the 
$16^4$ lattice and
from standard (solid line)
and boosted (dashed line) LPTH.  } \protect\label{fig:ggcomp}
\end{figure}

Finally, we can perform a cross-check of our perturbative calculations 
on the lattice and in the continuum as follows.
It is known~\cite{???} that at the one-loop level
\beq
\frac{\Lams}{\Lamlatt}=28.8.
\label{eq:ginv}
\eeq
This is a gauge-invariant quantity, since $\Lams$ and $\Lamlatt$ are
both gauge independent. 
We can insert in the identity
\beq
\Bigl(\frac{\Lams}{\Lam(\lambda)}\Bigr)
\Bigl(\frac{\Lam(\lambda)}{\Lamlatt}\Bigr)
= \frac{\Lams}{\Lamlatt}
\eeq
our
continuum result for $\Lams / \Lam$ and the lattice one for
$\Lam / \Lamlatt$, which we have evaluated for a generic covariant
gauge. For any value of $\lambda$ we get indeed 
the value (\eq{eq:ginv}) found in the literature.
\par
\vskip 0.5truecm

\section{Conclusions}
\label{section:conclusions}
We have shown in the pure gauge theory that a non-perturbative
determination of the QCD running coupling can be obtained from first
principles by a lattice study of the triple-gluon vertex.  We have
some evidence that systematic lattice effects are under control in our
calculation. The main features of our method are that LPTH is not
needed to match our results to $\MSB$ and that the extension to the
full theory does not present in principle any extra problem.  
Encouraged by our results in the quenched approximation,
we are now repeating our calculation with dynamical quarks.

\section{Acknowledgements} The numerical work was carried out
on the Connection Machine 200 at the University of Edinburgh, and on
the Fermilab lattice supercomputer, ACPMAPS.  B.~All\'es acknowledges
an Italian INFN postdoctoral fellowship.  D.~Henty, C.~Parrinello and
D.~Richards acknowledge the support of PPARC through a Personal
Fellowship (DH), Advanced Fellowships (CP and DGR), and grant GR/J
21347 (CP).  C.~Pittori acknowledges the support of an HCM Individual
Fellowship ER-BCHBICT930887.  This work was supported in part by the
DOE under contract DE-AC02-76CH03000.

We thank C.~Michael, O.~P\`ene and G.C.~Rossi for many helpful
suggestions and for reading the manuscript. We also thank
J.I.~Skullerud for some help with the numerical work.
\newpage

\section{Appendix}

We start by giving the details of the (quenched) one-loop
continuum perturbative 
calculation needed to relate the scale parameter $\Lambda$ in 
the $\widetilde{MOM}$ scheme to the one in $\MSB$. 

The three-gluon vertex in the $\MSB$ scheme, 
calculated at the asymmetric point and
at $p^2=\mu^2$,  
can be written in a generic covariant gauge as (we suppress colour indices)
\begin{eqnarray}
\Gamma^{(3)}_{\MSB \ \alpha \beta \gamma} (p / \mu,0,- p / \mu) 
\vert_{p^2 = \mu^2} & = & 
\left[ 1 + \frac{g^2}{16 \pi^2} C_V (\lambda) \right] 
\ \Gamma^{(3)}_{tree \ \alpha \beta \gamma} (p,0,-p) \nonumber \\
& + & i \frac{g^3}{16 \pi^2} 
C_{extra} (\lambda) \left[ \delta_{\alpha \gamma} p_{\beta} -
\frac{p_{\alpha} 
p_{\beta} p_{\gamma}}{p^2} \right],
\protect\label{eq:a1}
\end{eqnarray}
where $\lambda$ is the gauge parameter and  
\begin{eqnarray}
C_V (\lambda) & = & 
\frac{3}{2} \ \left[ \frac{61}{18} + \frac{1}{2} \lambda^2 \right], 
\nonumber \\
C_{extra} (\lambda) & = & 
\frac{3}{2} \ \left[ - \frac{37}{6} + 3 \lambda + \frac{1}{2} 
\lambda^2 \right]. 
\end{eqnarray}
The (Euclidean) tree-level vertex is
\begin{equation}
\Gamma^{(3)}_{tree \ \alpha \beta \gamma} (p,0,-p) = -i g \ 
\left[\delta_{\alpha \beta} p_{\gamma} + \delta_{\gamma \beta} p_{\alpha} 
- 2 \delta_{\alpha \gamma} p_{\beta} \right].
\end{equation}
Note that despite the vanishing of one of the external momenta,
the renormalised vertex (\protect\ref{eq:a1}) is a finite quantity, as the 
infrared behaviour is completely controlled by the off-shell gluon~\cite{BF}.

Setting $\alpha = \gamma$ and summing over $\alpha$ we get
\begin{eqnarray}
  &&{\sum_{\alpha} \Gamma^{(3)}_{\MSB \ \alpha \beta \alpha} 
(p / \mu,0,- p / \mu)
\vert_{p^2 = \mu^2}} = \nonumber \\
&& 6 i g p_{\beta} \left[ 1 + \frac{g^2}{16 \pi^2} 
\left(C_V (\lambda) + \frac{C_{extra} (\lambda)}{2}\right) \right].
\label{eq:comb}
\end{eqnarray}
Whereas in the $\MSB$ scheme only the pole part 
appears in the definition of renormalisation constants,
nontrivial finite terms are included in momentum subtraction schemes.
In particular, in our $\widetilde{MOM}$ scheme the finite
combination $(C_V + C_{extra}/2)$, which appears in eq.~(\ref{eq:comb}),
enters in the definition (\ref{eq:baba}) of $Z_V$.
Hence, by computing
\beq
g^2(\mu) 
= Z_A^{3}(\epsilon,\mu) Z_{V}^{-2} (\epsilon,\mu) \ g^2_{0}(\epsilon)
\label{eq:renvert2}
\eeq
in the two schemes, one finds that
the coefficient $C_{\MSB,\widetilde{MOM}}$, which relates 
the two couplings according to 
 (\protect\ref{eq:gengen}), is given by 
the expression
\begin{equation}
C_{\MSB,\widetilde{MOM}} = - \frac{1}{4 \pi} \left[3 C_A (\lambda) - 2 
\left( C_V (\lambda) + \frac{C_{extra} (\lambda)}{2}\right) \right],
\end{equation} 
where
\begin{equation}
C_A (\lambda) = 3 \ \left[ \frac{97}{36} + \frac{1}{2} \lambda +
\frac{1}{4} 
\lambda^2 \right]
\end{equation}
is the finite contribution coming from the gluon self-energy at one-loop~\cite{Cel}. 
By setting $\lambda=0$, we obtain the result in the Landau gauge
\begin{equation}
C_{\MSB,\widetilde{MOM}} = - 1.856807669 \ldots
\end{equation}
Finally, the ratio (\ref{eq:matchcont}) of the $\Lambda$ parameters  
is obtained from 
\begin{equation}
\frac{\Lambda^{(0)}_{\MSB}}{\Lambda^{(0)}_{\widetilde{MOM}}} = 
\exp \left[ \frac{C_{\MSB,\widetilde{MOM}}}{8 \pi \ b_0} \right] = 0.35,
\end{equation}
where $b_0 = \frac{11}{16 \pi^2}$ in the quenched approximation.

Turning now to our lattice calculations, we observe that the part of 
the one-loop three-gluon vertex on the lattice that is 
proportional to the 
tree-level vertex can be written as
\begin{eqnarray}
 \Gamma^{(3)}_{L} & = &
 \Gamma^{(3)}_{\rm tree} \times
\left( 1 + g^2_0 A_{\rm L} \right) \nonumber \\
& = &\Gamma^{(3)}_{\rm tree} \times
\left(1 + g_0^2 A_{\MSB} + g_0^2 C_{\rm L} \right),
\end{eqnarray}
\noindent 
where $A_{\rm L}$ and $A_{\MSB}$ stand for the (momentum dependent)
lattice
and continuum ($\MSB$ scheme) contributions respectively.
The quantity $C_{\rm L}$, which relates the pure lattice result to the
$\MSB$ scheme, is momentum independent.
Extra terms not proportional to the tree-level vertex are equal on the 
lattice and in the continuum, thus they do not contribute to $C_{\rm L}$, 
which can be computed by evaluating the diagrams of Figure~6.

\begin{figure}[t]
\vspace{-1cm}
\centerline{
\ewxy{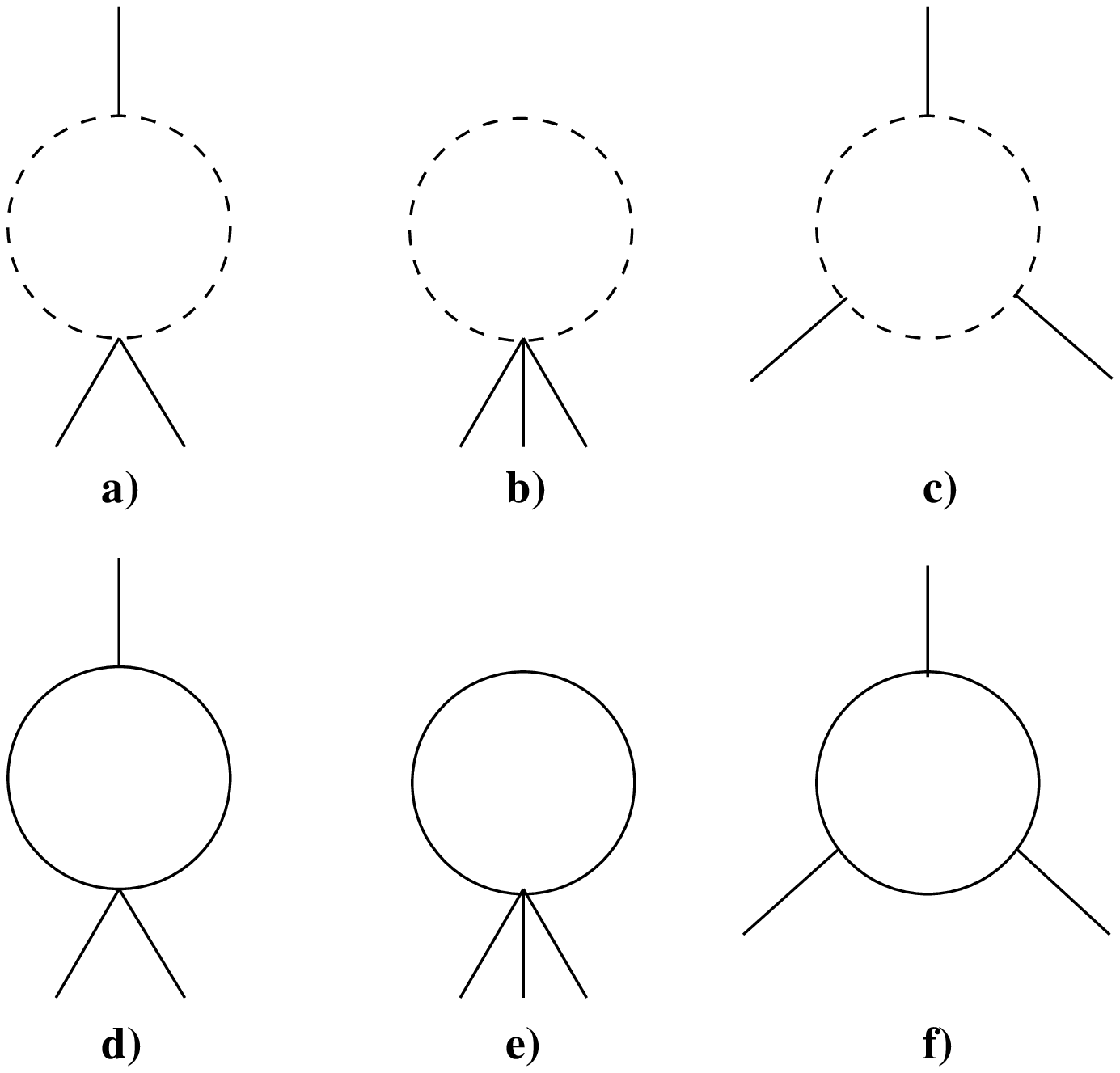}{256pt}}
\caption{Diagrams for the one-loop
three-gluon vertex on the lattice.
Dashed and solid lines indicate
ghosts and gluons respectively.
}
\protect\label{fig:alles}
\end{figure}

The contribution of each one to $C_{\rm L}$ is
\begin{eqnarray}
\hbox{a)} & =& 0 \nonumber \\
\hbox{b)} & =& 0 \nonumber \\
\hbox{c)} & =& {N \over 24} \left( {1 \over {16 \pi^2}} L
- {1 \over 4} J + {1 \over 6} Z_0 \right) \nonumber \\
\hbox{d)} & = & {N \over 4}
\left( {9 \over {16 \pi^2}} L
- {9 \over 4} J - {3 \over {16}} +
{{143} \over {48}} Z_0 \right) + \nonumber \\
&&  (1-\lambda) {N \over {8}}
\left( -{{3} \over {16 \pi^2}} L + 
{3 \over {4}} J - {{13} \over {8}} Z_0 \right) \nonumber \\
\hbox{e)} & = &{1 \over {8N}} - N \left( {1 \over {64}} +
{5 \over 8} Z_0 \right)
+ (1 - \lambda) N {{11 } \over {96}}  Z_0 \nonumber \\
\hbox{f)} & = &{N \over {8}} \left( -{{13} \over {16 \pi^2}}
L + {{13} \over 4} J - {13 \over 8} Z_0 +
{16 \over {96 \pi^2}} \right) + \nonumber \\
&&  (1 - \lambda) {N \over {8}}
\left( {9 \over {16 \pi^2}} L -
{9 \over {4}} J + {{23} \over {24}} Z_0 \right).
\label{eq:daaafres}
\end{eqnarray}
\noindent 
In these equations $L$, $Z_0$ and $J$ are~\cite{kawai,gonzalez}

\begin{eqnarray}
L & \equiv & \ln \mu^2 a^2 + \gamma_{\rm euler} - \ln 4 \pi \nonumber \\
Z_0 & \equiv & \int^{+\pi}_{-\pi} { {\hbox{d}^4q} \over {(2 \pi)^4}}
{1 \over {\hat{q}^2}} = 0.1549333902311, \nonumber \\
J & \equiv & 0.0465621749414. 
\end{eqnarray}

\noindent Finally $\mu$ is the mass scale introduced during the
dimensional regularisation of the continuum $\MSB$ scheme.
These formulae are
valid for the gauge group $SU(N)$. Notice that both $A_{\rm L}$ and
$A_{\MSB}$ depend on $\lambda$, $\lambda^2$ and $\lambda^3$, while
$C_{\rm L}$ depends only on $\lambda$.

Collecting all contributions in eq.~(\ref{eq:daaafres}), we obtain the
result for $C_{\rm L}$

\begin{eqnarray}
 C_{\rm L} &= & {1 \over {8 N}} + {2 \over 3} N
\left( {1 \over {16 \pi^2}} L - {1 \over 4} J -
{11 \over 96} Z_0 - {3 \over {32}} + {1 \over {32 \pi^2}} \right) + 
\nonumber \\
&& (1 - \lambda) {3 \over 4} N \left(
{1 \over {16 \pi^2}} L - {1 \over 4} J +
{1 \over {24}} Z_0 \right). 
\end{eqnarray}

Now, using the results of reference~\cite{BF}
we get the result for $A_L$ in the
Landau gauge
\begin{equation}
A_L = {17 \over 4} {1 \over {16 \pi^2}} \ln p^2 a^2 - 0.294728.
\end{equation}

\end{document}